\documentclass[twoside]{dis07}
\usepackage[latin1]{inputenc}
\usepackage[dvips]{graphicx,epsfig,color}
\usepackage{wrapfig,rotating}
\usepackage{amssymb,amsmath,array}

\begin{document}
\title{Parton Distributions for the LHC}

\author{R.S. Thorne$^{1}$\thanks{Royal Society University Research Fellow}, A.D. Martin$^2$, W.J. Stirling$^2$ and G. Watt$^1$
%
\vspace{.3cm}\\
%
1- Department of Physics and Astronomy, University College London, WC1E 6BT, UK
\vspace{.1cm}\\
2- Institute for Particle Physics Phenomenology, University of Durham, DH1 3LE, UK \\
}

\maketitle

\begin{abstract}
  We present a preliminary set of updated NLO parton distributions.  For the first time we have a quantitative extraction of the strange quark and antiquark distributions and their uncertainties determined from CCFR and NuTeV dimuon cross sections.  Additional jet data from HERA and the Tevatron improve our gluon extraction.  Lepton asymmetry data and neutrino structure functions improve the flavour separation, particularly constraining the down quark valence distribution.
\end{abstract}

There are many reasons why an update \cite{url} to the MRST2004 parton distributions \cite{MRST04} is necessary.  The MRST (now MSTW) group have used a general-mass variable flavour number scheme (VFNS) since 1998 \cite{MRST98}, but a new scheme \cite{nnlovfns}, although primarily devised for use at NNLO \cite{MRST06}, also changes the details of the NLO implementation.  We now make use of the \texttt{fastNLO} package \cite{fastNLO}, which provides an efficient combination of NLO partonic cross sections with parton distributions, allowing for a more rigorous inclusion of jet data into global parton fits.

\begin{wrapfigure}{r}{0.5\columnwidth}
\centerline{\includegraphics[width=0.5\columnwidth]{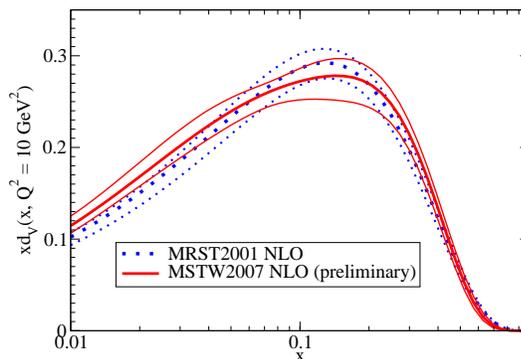}}
\caption{The distribution $xd_V(x,10\,{\rm GeV}^2)$.}\label{Fig:dval}
\end{wrapfigure}

There are also many new data sets available to include: CHORUS and NuTeV neutrino structure functions $F_2^{\nu,\bar\nu}(x,Q^2)$ and $F_3^{\nu,\bar\nu}(x,Q^2)$ \cite{nuStruct}, CCFR and NuTeV dimuon cross sections \cite{NuTeVdimuon} providing a direct constraint on $s$ and $\bar{s}$, CDF Run II lepton asymmetry data \cite{newleptasym} in two different $E_T$ bins, HERA jet data \cite{HERAjets}, and new CDF Run II high-$E_T$ jet data \cite{newcdf}.  We also include all recent charm and bottom HERA structure function data \cite{newcharm}.

Let us consider the effects of some of these new data sets.  It has already been noticed that the NuTeV neutrino structure functions are not completely compatible with the older CCFR data, both by the experiment themselves and by groups performing fits \cite{DIS06,CTEQnu}.  Previous parton distributions were perfectly compatible with the CCFR data when using EMC-inspired $Q^2$-independent nuclear correction factors, whereas the NuTeV structure function data are difficult to fit at high $x$.  Moreover, recent CHORUS structure function data (which were taken with a lead rather than an iron target) turn out to be fairly consistent with the CCFR data.  Thus, there is a problem at high $x$.  However, the partons in this region are already very well determined from charged lepton structure functions.  The important information from charged current data is in the region $x<0.3$, probing the separation into valence quarks and sea quarks, and there is general consistency between data sets here.  Hence, we choose to exclude from the fit all neutrino structure function data for $x>0.5$.  In our analysis we now also implement more sophisticated flavour-dependent nuclear correction factors \cite{Nuccorr} extracted using NLO partons, allowing a $\sim 3\%$ uncertainty on these factors.

\begin{figure}
  \centerline{\includegraphics[width=0.5\columnwidth]{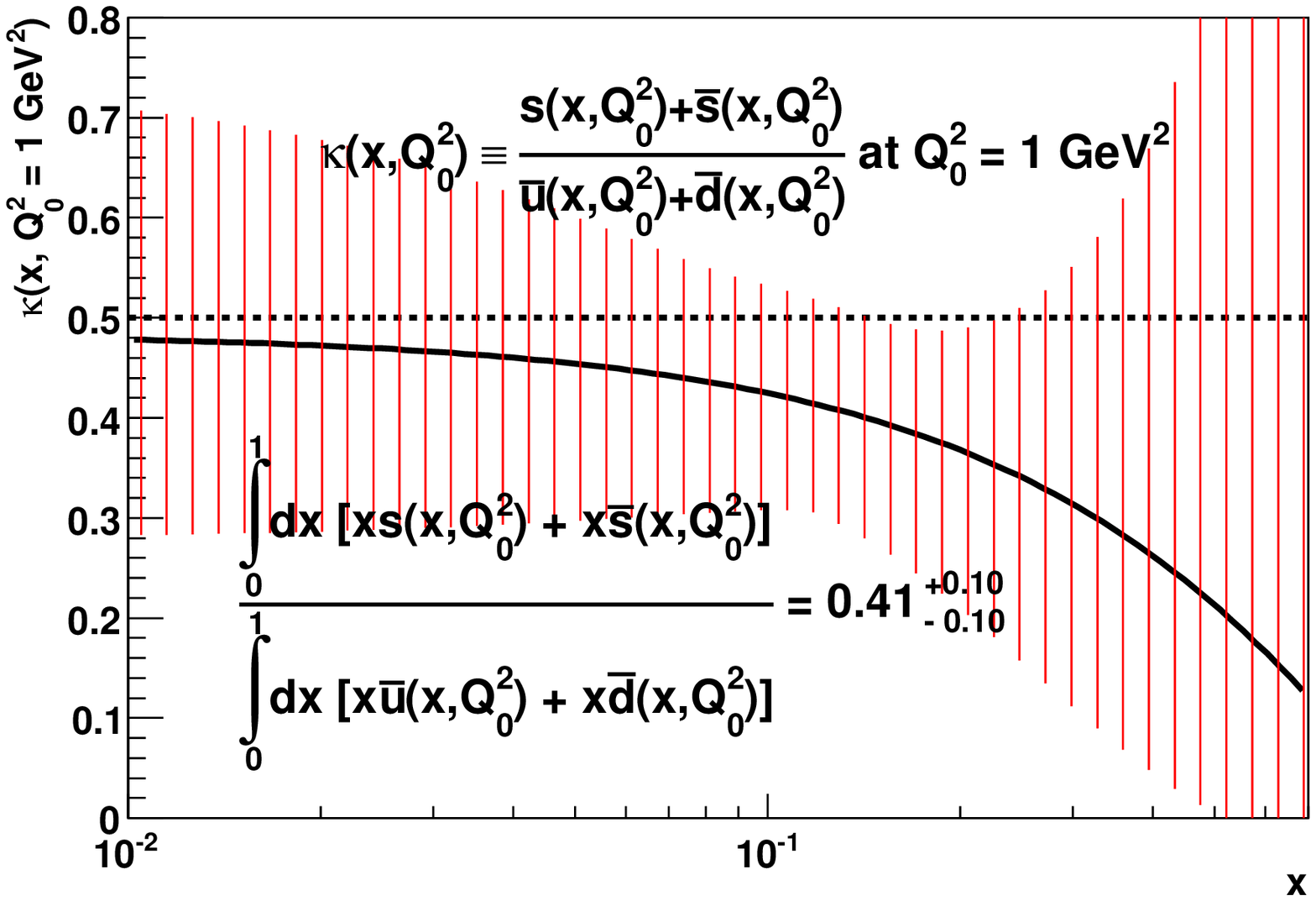}
    \hspace{-0.5cm}\includegraphics[width=0.5\columnwidth]{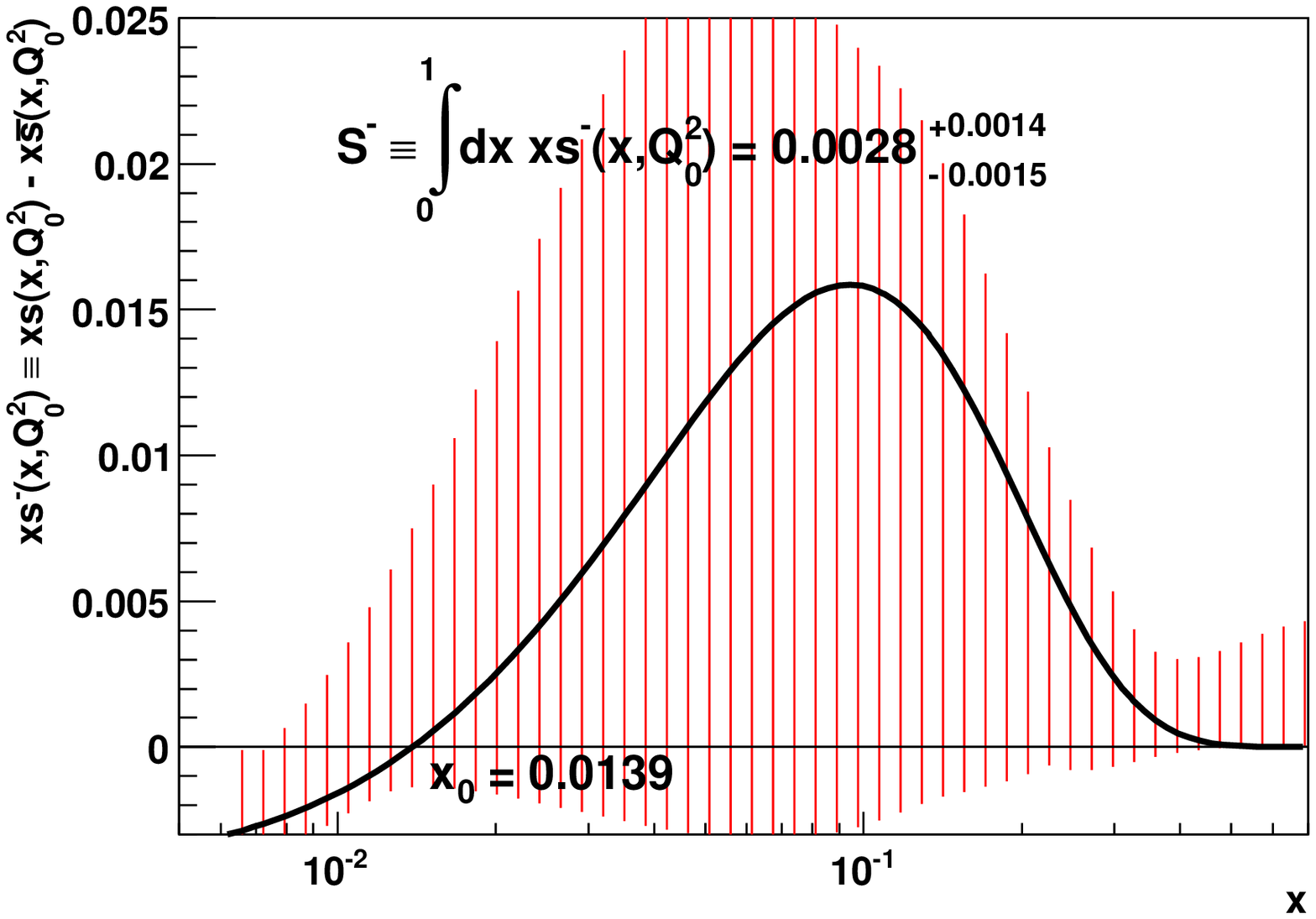}}
  \caption{The input distributions $s^+(x,Q_0^2)$ (left) and $s^-(x,Q_0^2)$ (right).}\label{Fig:splus}
\end{figure}
The CDF lepton asymmetry data constrain similar partons as the neutrino structure functions.  The $W$ asymmetry at the Tevatron is defined by $A_W(y_W) \propto {\rm d}\sigma(W^+)/{\rm d}y_W-{\rm d}\sigma(W^-)/{\rm d}y_W \approx u(x_1)d(x_2)-d(x_1)u(x_2)$, where $x_{1,2}=(M_W/\sqrt{s})\exp(\pm y_W)$.  In practice it is the final state leptons from the $W$ decay that are detected, so it is really the lepton asymmetry $A(y_l)= [\sigma(l^+)-\sigma(l^-)]/[\sigma(l^+)+\sigma(l^-)]$ which is measured.  Defining the emission angle of the lepton relative to the proton beam in the $W$ rest frame by $\cos^2 \theta^* = 1 - 4E_T^2/M_W^2$ leads to $y_{l}= y_W \pm 1/2\ln[(1+\cos\theta^*)/(1-\cos \theta^*)]$.  To a good approximation 
$$
\sigma(l^+)-\sigma(l^-) \propto u(x_1)d(x_2)(1-\cos\theta^*)^2 + \bar{d}(x_1)\bar{u}(x_2)(1+\cos\theta^*)^2 - d(x_1)u(x_2)(1+\cos\theta^*)^2,
$$
where the antiquark term is boosted by the large $(1+\cos\theta^*)^2$ and the asymmetry is fairly sensitive to antiquarks at lower $E_T$.  Hence, it probes the separation into valence and sea quarks, particularly for the less well-constrained down quark.  We find that the CDF Run II data influence $d(x,Q^2)$, and that there is some tension with the neutrino structure function data.  Our fit gives $\chi^2 = 41$ for the $22$ lepton asymmetry data points, although much of this $\chi^2$ comes from only 2 data points.  Overall, $d_V(x,Q^2)$ has a slightly different shape to previously \cite{MRSTerror1}, seen in Fig.~\ref{Fig:dval}.  The uncertainty is generally bigger than before, despite more constraints, due to a better parameterisation when determining uncertainty eigenvectors.

The CCFR and NuTeV dimuon cross sections for neutrino (antineutrino) scattering, $\nu_{\mu}(\bar{\nu}_\mu)N\to\mu^+\mu^-X$, are sensitive to the strange quark (antiquark) distribution through the LO partonic process $\nu_{\mu}s(\bar{\nu}_\mu\bar{s})\to c\mu^-(\bar{c}\mu^+)X$.  In previous MRST fits, CCFR dimuon data have been indirectly used to justify the input parameterisation assumption of
$$
s(x,Q_0^2) = \bar s(x,Q_0^2) = \frac{\kappa}{2}[\bar u(x,Q_0^2) + \bar d (x,Q_0^2)]\qquad (\kappa \approx 0.5),
$$
at the input scale of $Q_0^2=1\,{\rm GeV}^2$.  We now relax this assumption and fit the CCFR/NuTeV dimuon cross sections directly.  Defining $s^+\equiv s+\bar{s}$ and $s^-\equiv s-\bar{s}$, then the input forms are taken to be $s^+(x,Q_0^2) = A_+\,(1-x)^{\eta_+}\,S(x,Q_0^2)$ and $s^-(x,Q_0^2) = A_-\,x^{-1+\delta_-}\,(1-x)^{\eta_-}\,(1-x/x_0)$, where $S(x,Q_0^2)$ is the total sea quark distribution and $x_0$ is determined by zero strangeness of the proton, i.e.~$\int_0^1\!{\rm d}x\;s^-(x,Q_0^2) = 0$.  The preference for extra freedom in both $s^+$ and $s^-$ is confirmed by the fit, with the global $\chi^2$ improving by 15 when $s^+$ is allowed to go free, and by a further 14 when $s^-$ is allowed to go free.  There is no improvement with the addition of further free parameters.  Indeed, $A_-$ and $\delta_-$ are highly correlated so we fix $\delta_-$ at 0.2, i.e.~a valence-like value.

\begin{wrapfigure}{r}{0.55\columnwidth}
  \centerline{\includegraphics[width=0.55\columnwidth]{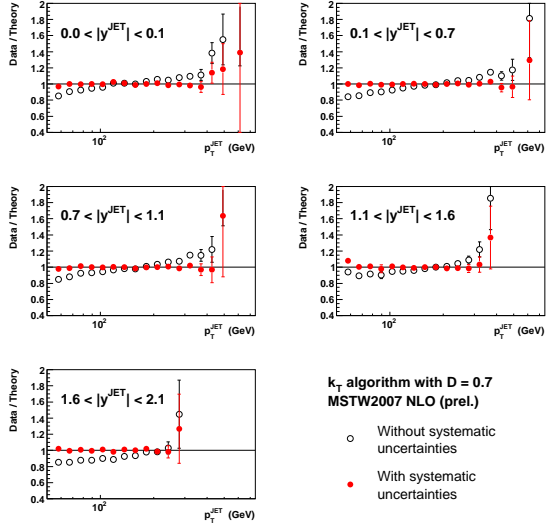}}
  \caption{The fit to the CDF Run II jet data.}\label{Fig:cdf}
\end{wrapfigure}

The input $s^+$ and $s^-$ distributions from the best fit, with approximately 90\% confidence level uncertainty bands, are shown in Fig.~\ref{Fig:splus}.  There is a reduced ratio of strange to non-strange sea distributions compared to our previous default $\kappa=0.5$, and a marked suppression at high $x$, i.e.~low $W^2$, probably due to the effect of the strange quark mass.  The integrated momentum asymmetry is positive, $0.0028^{+0.0014}_{-0.0015}$ at $Q_0^2 = 1\,{\rm GeV}^2$, decreasing to $0.0021^{+0.0010}_{-0.0011}$ at $Q^2 = 10\,{\rm GeV}^2$.  The results on both $s^+$ and $s^-$ are qualitatively consistent with those obtained by the CTEQ group \cite{CTEQstrange}.  Directly fitting the $s$ and $\bar{s}$ distributions affects the uncertainties on the light quarks.  Until recently all parton fitting groups assumed $s^+$ to be a fixed proportion of the total sea in the global fit.  An independent uncertainty on $s$ and $\bar{s}$ feeds into that on the $\bar{u}$ and $\bar{d}$ quarks, since the neutral current DIS data on $F_2(x,Q^2)$ constrain the combination $4/9 (u+\bar{u})+1/9(d+\bar{d}+s+\bar{s})$.  Consequently, the size of the uncertainty on the sea quarks at $x\sim 0.001-0.01$ at hard scales $Q^2\sim M_W^2$ roughly doubles from $\sim 1.5\%$ to $\sim 3\%$. 

\begin{wrapfigure}{r}{0.5\columnwidth}
\centerline{\includegraphics[width=0.5\columnwidth]{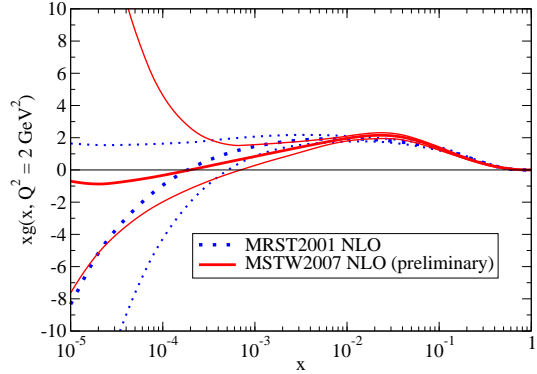}}
\caption{The distribution $xg(x,2\,{\rm GeV}^2)$.}\label{Fig:gluon}
\vspace{-0.5cm}
\end{wrapfigure}

We now use the \texttt{fastNLO} \cite{fastNLO} package during the fit to calculate Tevatron and HERA jet cross sections at NLO, improving upon previous approximations for the former and allowing our first analysis of the latter.  We find a slight change in the shape of the gluon distribution using only Tevatron Run I data (for which we now include previously absent hadronisation corrections).  The fit to the HERA jet data is excellent, although we do not find that they provide a strong constraint within the global fit.  We also include CDF Run II inclusive jet data \cite{newcdf} obtained in different rapidity bins using the $k_T$ jet algorithm.  We obtain a very good fit, $\chi^2=56$ for $76$ data points, as seen in Fig.~\ref{Fig:cdf}, at the expense of a slight deterioration in the fit to the D{\O} Run I inclusive jet data \cite{D0jets}.  The CDF Run II data prefer a smaller gluon distribution at very high $x$ compared to the Run I data included in the MRST2004 fit \cite{MRST04}.  The gluon distribution is shown in Fig.~\ref{Fig:gluon} compared to MRST2001 \cite{MRSTerror1}.  The uncertainty is similar at high $x$, but is actually larger overall due to allowing an additional parameter to go free when determining the eigenvectors.  The uncertainty on the gluon is extremely large at $x=10^{-5}$ due to our lack of theoretical prejudice on the form taken in this region.

\begin{wraptable}{r}{0.47\columnwidth}
  \centerline{\begin{tabular}{|c|c|c|c|}
      \hline
      $m_c$ &  $\chi^2_{\rm global}$  &  $\chi^2_{F_2^c}$  &  $\alpha_s(M_Z^2)$ \\
      (GeV) &  2659 pts.  &  78 pts. &  \\
      \hline
      1.2  & 2541  &  179  & 0.1183   \\
      1.3  & 2485  &  129  & 0.1191   \\
      1.4  & 2472  &  100  & 0.1206   \\
      1.5  & 2479  &  95  & 0.1213   \\
      1.6  & 2518  &  101  & 0.1223   \\
      1.7  & 2576  &  123  & 0.1221   \\
      \hline
  \end{tabular}}
  \caption{The dependence of the fit on $m_c$.}
  \label{Tab:table}
\end{wraptable}

Finally, we investigate the dependence of the fit on $m_c$, the pole mass of the charm quark.  We take the pole mass of the bottom quark to be $m_b = 4.75\,{\rm GeV}$.  We vary $m_c$ in steps of $0.1\,{\rm GeV}$, finding a preference for $m_c=1.4\,{\rm GeV}$ with an uncertainty $\sim 0.2\,{\rm GeV}$.  The results are shown in Table \ref{Tab:table}.  There is a clear correlation between $m_c$ and $\alpha_S(M_Z^2)$.  The value of the (NLO, $\overline{\rm MS}$) coupling for the best fit is $\alpha_S(M_Z^2) = 0.1206$, very similar to that in MRST2004 \cite{MRST04}.

Overall, the inclusion of new data and the changes in the analysis have a significant impact on our NLO parton distributions.  Based on the developments in this paper we will soon have fully updated NLO and NNLO partons for the LHC complete with uncertainties.

\begin{footnotesize} 

\end{footnotesize}

\end{document}